\documentclass[12pt]{article}
\usepackage{amssymb, latexsym,amsmath,natbib,graphicx,epstopdf,dsfont,outlines,tikz,placeins,amsthm,subfigure,relsize,sidecap}
\usepackage[margin=.95in]{geometry}
\usepackage[compact]{titlesec}
\newtheorem{theorem}{Theorem}[section]

\newtheorem{proposition}[theorem]{Proposition}

\bibpunct{(}{)}{;}{a}{}{,}

\newcommand{\bx}{\boldsymbol{X}}

\newcommand{\bPsi}{\boldsymbol\Psi}

\newcommand{\cx}{{\cal X}}

\begin{document}
\textwidth 6.5in

\begin{center}
{\Large\bf Latent Space Models for Dynamic Networks}

\vspace{0.5in}

Daniel K. Sewell and Yuguo Chen
\footnote{Daniel K. Sewell is Ph.D Candidate, Department of Statistics,
University of Illinois at Urbana-Champaign, Champaign, IL 61820 (E-mail: {\it dsewell2@illinois.edu}). Yuguo Chen
is Associate Professor, Department of Statistics, University of Illinois at
Urbana-Champaign, Champaign, IL 61820 (E-mail: {\it yuguo@illinois.edu}).
This work was supported in part by National Science Foundation grant DMS-11-06796.  The authors thank the editor, the associate editor, and a referee for valuable suggestions.}

\end{center}

\begin{abstract}
Dynamic networks are used in a variety of fields to represent the structure and evolution of the relationships between entities.  We present a model which embeds longitudinal network data as trajectories in a latent Euclidean space.  A Markov chain Monte Carlo algorithm is proposed to estimate the model parameters and latent positions of the actors in the network.  The model yields meaningful visualization of dynamic networks, giving the researcher insight into the evolution and the structure, both local and global, of the network.  The model handles directed or undirected edges, easily handles missing edges, and lends itself well to predicting future edges.  Further, a novel approach is given to detect and visualize an attracting influence between actors using only the edge information.
We use the case-control likelihood approximation to speed up the estimation algorithm, modifying it slightly to account for missing data.  We apply the latent space model to data collected from a Dutch classroom, and a cosponsorship network collected on members of the U.S. House of Representatives, illustrating the usefulness of the model by making insights into the networks.
\vspace{ 2mm}

\end{abstract}

\newpage

\section{INTRODUCTION}
Network analysis, and in particular dynamic network analysis, is a ubiquitous area of study, used by scientists in many distinct fields \citep{vivar2011models}.  Often studied are dynamic social networks, which come in a wide variety of forms (see the Special Issues on Network Dynamics in {\it Social Networks}, January 2010 and July 2012).  In this paper we consider data that come in the form of a set of actors and a sequence of sets of edges, each edge set having been measured at one of multiple time points.  Analyzing dynamic social networks is key to seeing how friendships form or dissolve, how politicians form loyalties or break ranks with their parties, how co-authorship patterns develop and change over time, etc.  Dynamic networks are also analyzed in epidemiological contexts \citep{bansal2010dynamic}, in analyzing terrorist networks \citep{carley2006dynamic}, and much more.

There exist numerous methods of modeling network data within a statistical framework \citep[for a survey on statistical network models, see][]{goldenberg2010survey}.  Some of these models are intended for static networks but have generative processes which can be thought of as dynamic, in the sense of building up the graph over a series of time points.  Examples of this notion can be found in the rewiring of ``small-world" networks \citep{watts1998collective}, the subsequent addition of edges in an Erd\"{o}s-R\'{e}nyi random graph model \citep{durrett2007random}, or the addition of actors and edges in a duplication-attachment model \citep{kumar2000stochastic}.  Other methods were developed for static networks and were then extended for the dynamic case.  One of the most well known methods of analyzing static networks is the exponential random graph model (ERGM) developed by \cite{frank1986markov}, and much attention is still being given to this class of models \citep[see, e.g., ][]{robins2007recent,bollobas2007phase}.  This was extended to analyzing networks observed over discrete time intervals by \cite{hanneke2010discrete} in the introduction of the temporal ERGM, or TERGM.  Using continuous time Markov processes, \cite{snijders1996stochastic} began a series of works corresponding to what is known as stochastic actor-oriented models.  Both of these last two approaches focus on the use of common network structures or user-defined objective functions.  The last commonly used approach to modeling networks that we will mention is the latent space model.  Latent space approaches aim to embed network information into some (usually low dimensional) latent space.  Benefits of using such an approach is that both local and global structures are modeled, transitivity is inherently incorporated in the model, meaningful visualizations are obtained, and the output is easily interpreted, lending itself to much qualitative inference.  While the bulk of the literature on latent space models is concerned with static networks, in this paper we will use this approach to model longitudinal network data.

The ideas behind latent space models have long been in use.  For example, \cite{nakao1993longitudinal} used multidimensional scaling to visualize and analyze the latent positions of the actors in Newcomb's fraternity data \citep{newcomb1956prediction}.  Two formal latent space models were introduced for static networks by \cite{hoff2002latent}, one of which placed the latent actor positions within a Euclidean space, the other placed the latent locations on a unit hypersphere while giving each actor an activity level.  This latter model was intended to allow for a lack of reciprocity in directed networks.  Estimation was performed using Markov chain Monte Carlo (MCMC), hence giving the full posterior of parameters and latent positions.  \citet{handcock2007model} expanded the Euclidean model of \citet{hoff2002latent} by allowing the latent space positions to follow a mixture of normals, hence allowing clustering to occur simultaneously with embedding in a Euclidean space.  \citet{krivitsky2009representing} expanded on this work by allowing asymmetrical edge probabilities.  \citet{schweinberger2003settings} used a similar approach as \citet{hoff2002latent} but used an ultrametric space rather than a Euclidean or hypersphere space to perform model-based clustering.  Further work was done in \citet{hoff2005bilinear}, where the author extended previous notions of ANOVA models of networks by including as interaction effects the hypersphere latent positions from \citet{hoff2002latent}.

A limited number of works has considered the temporal aspect of networks while implementing a latent space approach.  \citet{robinsondetecting1} presented a method of discovering change points in network behavior via using a $k$-dimensional simplex latent space.  \citet{foulds2011dynamic} developed a non-parametric infinite feature model, where the features are latent.  The work most related to our proposed approach is that of \citet{sarkar2005dynamic}, which extended the Euclidean latent space model of \citet{hoff2002latent} to dynamic networks, though only undirected networks can be analyzed with this method.  They developed a generalized multidimensional scaling (GMDS) to find the initial latent actor positions across discrete time points.  The authors then furthered this by using a conjugate-gradient method of optimizing an objective function.  While this is a speedy algorithm and hence can be used for larger data sets, the estimation is an ad hoc method which makes limited use of the available data.

In this paper, we propose a model which embeds dynamic directed, or undirected, network data into a latent Euclidean space, allowing each actor to have a temporal trajectory in this latent space.  Estimation of the model parameters and latent actor positions occur within a Bayesian framework using MCMC.  By using our approach, the user can observe much more easily how the network evolves over time, gain insight into global and local structures, handle missing data, make future predictions, and can detect the attracting influence one actor has on another actor's friendships (a concept we call edge attraction, which will be discussed later).  To improve the speed of the MCMC algorithm for large networks, we describe an approximation method which reduces the computational cost.

The remainder of the paper is organized as follows:  Section \ref{DynamicLatentSpaceModel} describes the proposed model for dynamic networks.  Section \ref{Estimation} outlines the Bayesian estimation of the model parameters and latent actor positions, as well as addressing the issue of scalability.  Section \ref{MissingData} details how to handle missing data.  Section \ref{Prediction} describes how to obtain network predictions.  Section \ref{NodalInfluence} gives a method for detecting and visualizing edge attraction.  Section \ref{Simulations} shows simulation results. Section \ref{RealDataAnalyses} presents the results from analyzing data collected from a Dutch classroom as well as from analyzing cosponsorship data collected on members of the U.S. House of Representatives.  Section \ref{Discussion} provides a brief discussion.

\FloatBarrier
\section{DYNAMIC LATENT SPACE MODEL}
\label{DynamicLatentSpaceModel}
We assume that data come in the form of $({\cal N},\{{\cal E}_t: t\in {\cal T}\})$, where ${\cal N}$ is the set of all actors, and ${\cal E}_t$ is the set of edges at time $t$.  For simplicity let ${\cal T}= \{1,2,\ldots,T\}$.  For the majority of the paper it will also be assumed that ${\cal E}_t$ consists of directed edges.  The general idea of the latent space approach is that this time series of graphs can be represented as a state space model, with a latent state variable representing the actors as positions in a low dimensional Euclidean space.  The closer two actors are in this latent Euclidean space, the more likely they are to form an edge.  This low dimensional space can be thought of as a characteristic space where the distance between actors represents how similar they are \citep{hoff2002latent}, or as a social space where the distance between two actors corresponds to the strength of the relationship between the two.

The notation to be used throughout the rest of the paper is as follows: $n=|{\cal N}|$ is the number of actors.  For a latent space $\Re^p$,  ${\bf X}_{it}$ is the $p$ dimensional vector of the $i^{th}$ actor's latent position at time $t$, and ${\cal X}_t$ is the $n\times p$ matrix whose $i^{th}$ row is ${\bf X}_{it}$.  $Y_t=\{y_{ijt}\}$ is the adjacency matrix of the observed network at time $t$, and $y_{ijt}=1$ if there is an edge from actor $i$ to actor $j$ at time $t$ and 0 otherwise.

The latent actor positions are modeled by a Markov process with the initial distribution
\begin{equation}
\pi({\cal X}_1|\boldsymbol\psi)= \prod_{i=1}^n N({\bf X}_{i1}|{\bf 0},\tau^2I_p),
\end{equation}
and transition equation
\begin{equation}
\pi({\cal X}_t|{\cal X}_{t-1},\boldsymbol\psi)=\prod_{i=1}^n N({\bf X}_{it}|{\bf X}_{i(t-1)},\sigma^2I_p)
\label{transition}
\end{equation}
for $t=2,3,\ldots,T$, where $I_p$ is the $p\times p$ identity matrix, $N({\bf x}|\boldsymbol{ \mu},\Sigma)$ denotes the normal probability density function with mean $\boldsymbol\mu$ and covariance matrix $\Sigma$ evaluated at ${\bf x}$, and $\boldsymbol\psi$ is a vector of parameters which will be defined shortly.

\begin{figure}[h]
\centering
\includegraphics[scale=0.25]{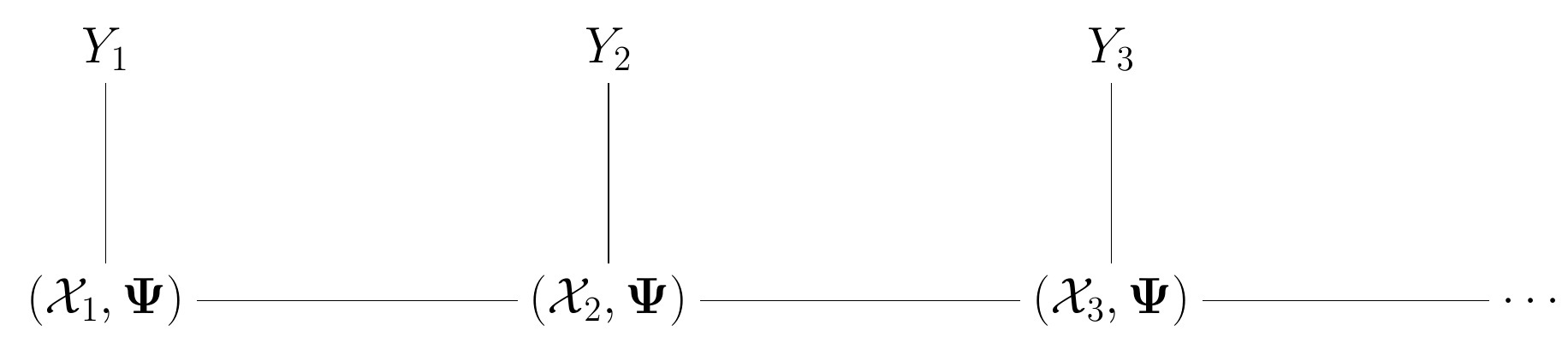}
\caption{Illustration of the dependence structure for the latent space model.  $Y_t$ is the observed graph, ${\cal X}_t$ is the unobserved latent actor positions, and $\boldsymbol\psi$ is the vector of model parameters.
}
\label{dep_struct}
\end{figure}

The observed networks at different time points are conditionally independent given the latent positions.  This dependence structure is illustrated in Figure \ref{dep_struct}.  Further, it is assumed that for any two (distinct) pairs $(i,j)$ and $(i',j')$, $y_{ijt}$ and $y_{i'j't}$ are independent conditioning on $({\cal X}_t,\boldsymbol\psi)$.  In formulating the observation equation of our model, we desire two main properties: first, the probability of an edge from actor $i$ to actor $j$ at time $t$ should increase as the distance between their latent positions decreases; second, the probability of an edge should depend on both who is sending and who is receiving the link, and we should further be able to determine the importance of each in edge formation; i.e., whether the identity of the sender or the identity of the receiver is more important in edge formation.  To this end, we use the formulation
\begin{equation}
\mathbb{P}(Y_t|{\cal X}_t,\boldsymbol\psi)=\prod_{i\neq j}\mathbb{P}(y_{ijt}=1|{\cal X}_t,\boldsymbol\psi)^{y_{ijt}}\cdot \mathbb{P}(y_{ijt}=0|{\cal X}_t,\boldsymbol\psi)^{1-y_{ijt}}=\prod_{i\neq j}\frac{\exp(y_{ijt}\eta_{ijt})}{1+\exp(\eta_{ijt})},
\label{obseq1}
\end{equation}
where
\begin{equation}
\eta_{ijt}:=\log \left( \frac{ \mathbb{P}(y_{ijt}=1|{\cal X}_t,\boldsymbol\psi)}{\mathbb{P}(y_{ijt}=0|{\cal X}_t,\boldsymbol\psi)}\right)= \beta_{IN}\left(1-\frac{d_{ijt}}{r_j}\right) + \beta_{OUT}\left(1-\frac{d_{ijt}}{r_i} \right),
\label{obseq2}
\end{equation}
and $d_{ijt}=\|{\bf X}_{it}-{\bf X}_{jt}\|$ and $\boldsymbol\psi = (\tau^2,\sigma^2,\beta_{IN},\beta_{OUT},r_{1:n})$ are the model parameters.  Here $r_{1:n}=(r_1,r_2,\ldots,r_n)$; similar notation will be used throughout the rest of the paper.   $\beta_{IN}$ and $\beta_{OUT}$ are global parameters which reflect the importance of popularity and social activity respectively.  The $r_i$'s are positive actor specific parameters that represent each actor's social reach and is reflective of the tendency to form and receive edges.  Within the latent space, there is also the geometrical interpretation of $r_i$ forming a radius around the $i^{th}$ actor, as we will see later.  For model identifiability,  the $r_i$'s are constrained so that $\sum_{i=1}^nr_i=1$. This parameterization emulates both the distance and projection models of \cite{hoff2002latent} for static networks, given as $\eta_{ij}=\beta(1-d_{ij})$ and $\eta_{ij}=\beta+{\bf X}_i'{\bf X}_j/\|{\bf X}_j\|$ respectively, by utilizing the visually appealing and intuitive Euclidean space for the latent positions while incorporating the individual actors' ``sociability," or social reach, while also accounting for both activity and popularity.

\cite{krivitsky2009representing} built onto Hoff et al.'s model by including additive random individual effects.  Here our parameterization links the actors' individual effects to the latent space, in the sense that the social reach dampens or augments the effect of the distance between the two actors, rather than having the individual actor effects be constant additive effects; thus these two parameterizations are in fact different, rather than being subsets of each other.  In some sense their model is more flexible in that an actor has both an indegree effect and an outdegree effect.  Our model can be trivially extended to account for this by simply allowing $r_{1:n}$ to be replaced by two sets of parameters $r_{1:n}^{(IN)}$ and $r_{1:n}^{(OUT)}$.  We applied this more complex model on the two real data sets presented in Section \ref{RealDataAnalyses} with no improvement in model fit.  Hence our focus remains on the simpler model, given in (\ref{obseq2}).

\begin{figure}[h]
\centering
\subfigure[]{
\includegraphics[scale=0.25]{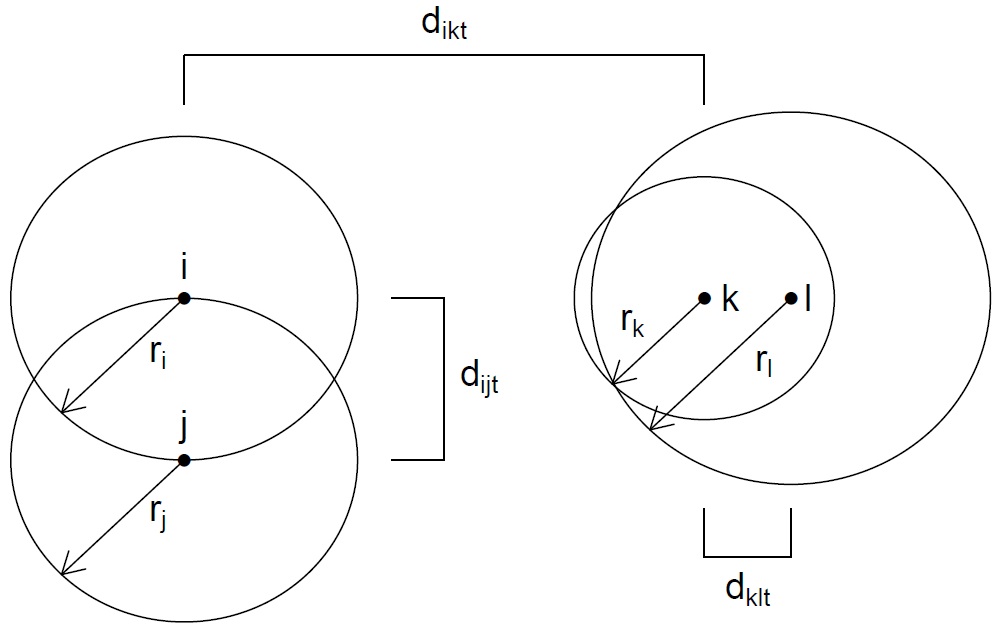}
\label{socialReach_fig}
}
\subfigure[]{
\includegraphics[scale=0.25]{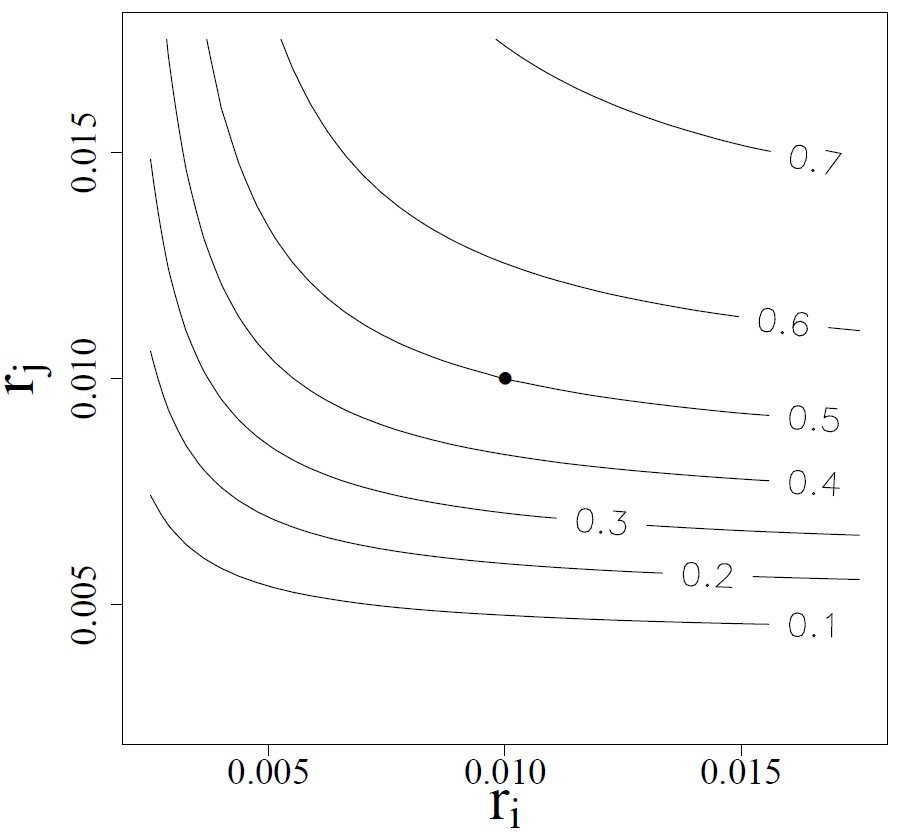}
\label{parmInterpretation_case1}
}
\caption{(a) Illustration of how to interpret social reach parameters $r_{1:n}$ in the case that $\beta_{IN},\beta_{OUT}>0$; the probability of an edge from $i$ to $k$ is less than 1/2, from $k$ to $l$ is greater than 1/2, and from $i$ to $j$ is equal to 1/2.  (b) Contour plot of $\mathbb{P}(y_{ijt}=1|{\cal X}_t,\boldsymbol\psi)$, where $\beta_{IN}=2$, $\beta_{OUT}=1/2$ and $d_{ijt}=0.01$.  The point $r_i=r_j=d_{ijt}$ is marked with a dot.}
\end{figure}

In the following discussion we make the (reasonable) assumption that both $\beta_{IN}>0$ and $\beta_{OUT}>0$ (the other possible case, discussed in the Supplementary Material, is where either $\beta_{IN}<0$ and $\beta_{OUT}>|\beta_{IN}|$ or $\beta_{OUT}<0$ and $\beta_{IN}>|\beta_{OUT}|$).  The interpretation of the radii that comes naturally from (\ref{obseq2}) is that $r_i$ marks the radius within the latent social space of the $i^{th}$ actor's social reach.  This is evident in that if the distance between two actors are within each other's radii, i.e., $d_{ijt}<\min(r_i,r_j)$, then the probability of an edge is greater than $1/2$; if they are outside each other's radii, i.e., $d_{ijt}>\max(r_i,r_j)$, then the probability of an edge is less than $1/2$; and if the distance between the two actors equals both radii, i.e., $d_{ijt}=r_i=r_j$, then the probability of an edge equals $1/2$.  These scenarios are illustrated in Figure \ref{socialReach_fig}.  Thus a larger radius implies an increasing propensity to send and receive ties.  This fact is further illustrated in Figure \ref{parmInterpretation_case1}, where the probability $\mathbb{P}(y_{ijt}=1|{\cal X}_t,\boldsymbol\psi)$ is shown in a contour plot, allowing $r_i$ and $r_j$ to vary, with distance $d_{ijt}=0.01$, $\beta_{IN}=2$ and $\beta_{OUT}=1/2$.

Concerning the global parameters $\beta_{IN}$ and $\beta_{OUT}$, if $\beta_{IN}>\beta_{OUT}$ ($\beta_{OUT}>\beta_{IN}$) then we can conclude that the probability of an edge from actor $i$ to actor $j$ (from actor $j$ to actor $i$) is determined more by the radius of $j$ than by the radius of $i$.  This is also illustrated in Figure \ref{parmInterpretation_case1}, where it is apparent that the probability of an edge from $i$ to $j$ increases much faster when we fix a value of $r_i$ and allow $r_j$ to increase than vice versa.  Thus if $\beta_{IN}>\beta_{OUT}$ then the edges of the network are determined more by the popularity of the actors than by their activity, i.e., the identity of the receiver of the edge is more important than the identity of the sender, and if $\beta_{OUT}>\beta_{IN}$ the edges of the network are determined more by the activity of the actors than by their popularity, i.e., the identity of the sender is more important than the identity of the receiver.

\section{ESTIMATION}
\label{Estimation}
\subsection{Posterior Sampling}
\label{PosteriorSampling}
We adopt a Bayesian approach, and hence we wish to make inferences based on $\pi({\cal X}_{1:T},\boldsymbol\psi|Y_{1:T})$, where $\boldsymbol\psi=(\tau^2,\sigma^2,\beta_{IN},\beta_{OUT},r_{1:n})$.  We implement a Metropolis-Hastings (MH) within Gibbs MCMC scheme as suggested by \cite{geweke2001bayesian} to sample from the posterior, thus giving point estimates and uncertainties.  We set the priors on the parameters as follows: assume that $\beta_{IN}\sim N(\nu_{IN},\xi_{IN})$, $\beta_{OUT}\sim N(\nu_{OUT},\xi_{OUT})$, $\sigma^2\sim IG(\theta_{\sigma},\phi_{\sigma})$, $\tau^2\sim IG(\theta_{\tau},\phi_{\tau})$  and $ (r_1,r_2,\ldots,r_n)\sim\mbox{Dirichlet}(\alpha_1,\alpha_2,\ldots,\alpha_n)$, where $IG$ is the inverse gamma distribution.  The inverse gamma priors were chosen to be conjugate, and the Dirichlet prior is a natural selection for such constrained parameters.

The number of MCMC iterations required to reach convergence can be greatly reduced by appropriate initial values of the latent positions and model parameters.  We give a discussion and suggest initialization strategies in the Supplementary Material.

To sample via Metropolis-Hastings within Gibbs algorithm, we draw from the full conditional distributions iteratively.  These conditional distributions are either known in closed form or up to a normalizing constant and are given in the Supplementary Material.  The posterior sampling algorithm is
\begin{description}
\item[0.]Set the initial values of $({\cal X}_{1:T},\boldsymbol\psi)$ (e.g., to those described in the Supplementary Material).\vspace{-1pc}
\item[1.]  For $t=1,\ldots,T$ and for $i=1,\ldots,n$, draw ${\bf X}_{it}$ via MH using a normal random walk proposal. \vspace{-1pc}
\item[2.] Draw $\tau^2$ from its full conditional inverse gamma distribution. \vspace{-1pc}
\item[3.] Draw $\sigma^2$ from its full conditional inverse gamma distribution. \vspace{-1pc}
\item[4.] Draw $\beta_{IN}$ via MH using a normal random walk proposal. \vspace{-1pc}
\item[5.] Draw $\beta_{OUT}$ via MH using a normal random walk proposal. \vspace{-1pc}
\item[6.] Draw $r_{1:n}$ via MH using a Dirichlet proposal. \vspace{-1pc}
\item[] Repeat steps 1-6. \vspace{-1pc}
\end{description}

Due to the constraint on the radii ($\sum_{i=1}^nr_i=1$), it is necessary to, within the MH step, accept or reject all $n$ values simultaneously; hence it is important to keep the movements small, i.e., keep the means at the current values and the variance of the proposal small.  Therefore the proposal used to draw the new values $r_{1:n}^*$ is another Dirichlet distribution with parameters $(\kappa r_1, \kappa r_2, \ldots, \kappa r_n)$, where the $r_i$'s are the current values and $\kappa$ is some large constant.

One last note is that the posterior will be invariant to rotations, reflections, and translations of the latent positions.  Hence any inference must take into account the non-uniqueness of the estimates.  Similar to the approach described in \cite{hoff2002latent}, we perform a Procrustes transformation to reorient the sampled trajectories.  We set an $(nT)\times p$ reference trajectory matrix $\boldsymbol{{\cal X}}_0$, and after drawing new ${\bf X}_{it}$ for all $i$ and $t$, we construct from these new draws the new trajectory matrix $\boldsymbol{{\cal X}}=({\cal X}_1',\ldots,{\cal X}_T')'$.  In practice we used the initial latent positions to construct $\boldsymbol{{\cal X}}_0$.  The Procrustes transformation on $\boldsymbol{{\cal X}}$ using $\boldsymbol{{\cal X}}_0$ as the target matrix finds
${\mbox{argmin}}_{\boldsymbol{{\cal X}}^*}tr\left(\boldsymbol{{\cal X}}_0- \boldsymbol{{\cal X}}^* \right)' \left(\boldsymbol{{\cal X}}_0- \boldsymbol{{\cal X}}^* \right)$, where $\boldsymbol{{\cal X}}^*$ is some rotation of $\boldsymbol{{\cal X}}$; see, e.g., \cite{borg2005modern}.  By performing the Procrustes transformation on the trajectory matrix, we obtain a single rotation matrix $A$ with which we use to set  $\boldsymbol{{\cal X}}^{(\ell)}=\boldsymbol{{\cal X}}A$, where the superscript $^{(\ell)}$ denotes the stored values for the $\ell^{th}$ iteration; that is, we set ${\bf X}_{it}^{(\ell)}=A'{\bf X}_{it}$.  By so doing we are preserving the distances between any actors at any time points, i.e., $\|{\bf X}^{(\ell)}_{it}-{\bf X}^{(\ell)}_{js}\|=\|{\bf X}_{it}-{\bf X}_{js}\|$ for any actors $i$ and $j$ and any time points $t$ and $s$.

\subsection{Scalability}
\label{Scalability}
Scalability is an issue for latent space models for network data.  For static networks, this issue has been addressed through using variational Bayes \citep{salter2012variational} and also by using case-control principles from epidemiology \citep{raftery2012fast}.  This latter method reduced the computational cost (for static networks) of computing the log likelihood from $O(n^2)$ to $O(n)$.

The general strategy of the case-control log likelihood approximation is to write the log likelihood as two summations.  Assuming that the network becomes sparser as $n$ gets larger, the computational cost of the first of the two summations is linear with respect to $n$, and the cost of the second is quadratic.  The second summation is then replaced by a Monte Carlo estimate obtained from a subsequence of the actors, thus making the overall cost of computing the log likelihood linear in $n$.

This method as described by \cite{raftery2012fast}, however, cannot be directly extended to longitudinal network data containing missing edge values which is often the case, especially in social networks.  This is because all the links need to be known a priori.  By modifying how the log likelihood is decomposed into two summations, we can apply this same method without knowing all $y_{ijt}$ beforehand.  The details on this approximation are given in the Supplementary Material.

\FloatBarrier
\section{MISSING DATA}
\label{MissingData}
Missing data in social networks is not uncommon, and can come in various forms, such as boundary specification, non-response, and censoring by vertex degree \citep{kossinets2006effects}.  Here we specifically focus on non-responses, i.e., missing edge values.  For static networks there have been a number of methods proposed \citep[see, e.g., ][]{robins2004missing,huisman2009imputation}.  For dynamic networks, \cite{huisman2008treatment} compared several methods to handle missing edges in the context of a stochastic actor oriented model.  \cite{handcock2010modeling} developed a theoretical framework for networks in which only a subset of the dyads are observed; we use this framework in our discussion and refer the reader to Handcock and Gile's paper for more details.

Let ${\cal D}$ denote the sampling pattern; that is, ${\cal D}$ is the set of $n\times n$ matrices $\{D_1,\ldots,D_T\}$ where $D_{ijt}$ equals 1 if the dyad $y_{ijt}$ is observed and equals 0 otherwise.  Letting ${\cal Y}^{(obs)}$ and ${\cal Y}^{(mis)}$ denote the collection of observed edges and missing edges respectively, the complete data is $({\cal Y}^{(obs)},{\cal Y}^{(mis)}, {\cal D})$, and the incomplete (observed) data is $({\cal Y}^{(obs)}, {\cal D})$.  The unobserved edges ${\cal Y}^{(mis)}$ are considered missing completely at random (MCAR) if $\mathbb{P}({\cal D}|{\cal Y}^{(obs)},{\cal Y}^{(mis)},\xi)=\mathbb{P}({\cal D}|\xi)$, where $\xi$ is some set of parameters corresponding to the sampling pattern.  If, however, $\mathbb{P}({\cal D}|{\cal Y}^{(obs)},{\cal Y}^{(mis)},\xi)=\mathbb{P}({\cal D}|{\cal Y}^{(obs)},\xi)$, then the unobserved edges are considered missing at random (MAR).  The case where the pattern of unobserved edges depends on the unobserved edges themselves (called non-ignorable missing data) is a difficult scenario which is beyond the scope of this paper; thus we will continue the discussion assuming that the missing edges are either MCAR or MAR.

\cite{rubin1976inference} discussed weak conditions for which it is possible to ignore the process that causes missing data.  In our context, we are interested in the posterior distribution $\pi({\cal X}_{1:T},\boldsymbol\psi,{\cal Y}^{(mis)}|{\cal Y}^{(obs)},{\cal D})$; hence if the sampling pattern is ignorable, we may make inference based on the posterior distribution $\pi({\cal X}_{1:T},\boldsymbol\psi,{\cal Y}^{(mis)}|{\cal Y}^{(obs)})$, i.e., ${\cal X}_{1:T}$, $\boldsymbol\psi$ and ${\cal Y}^{(mis)}$ are independent of ${\cal D}$ given ${\cal Y}^{(obs)}$.  There are two sufficient conditions that must be satisfied in order for the sampling pattern to be ignorable \citep{rubin1976inference}.  First, the sampling pattern parameters $\xi$ are {\it a priori} independent with the data $({\cal Y}^{(mis)},{\cal Y}^{(obs)})$, latent positions ${\cal X}_{1:T}$ and model parameters $\boldsymbol\psi$, i.e., $\pi({\cal Y}^{(mis)},{\cal Y}^{(obs)},{\cal X}_{1:T},\boldsymbol\psi,\xi)=\pi({\cal Y}^{(mis)},{\cal Y}^{(obs)},{\cal X}_{1:T},\boldsymbol\psi)\pi(\xi)$.  Second, the space of $(\xi,{\cal X}_{1:T},\boldsymbol\psi)$ is a product space, i.e., if $\xi\in\Xi$, ${\cal X}_{1:T}\in\boldsymbol{{\cal X}}$ and $\boldsymbol\psi\in\boldsymbol\Psi$ then $(\xi,{\cal X}_{1:T},\boldsymbol\psi)\in\Xi\times\boldsymbol{{\cal X}}\times\boldsymbol\Psi$.  If these two conditions are met and the missing edges are either MCAR or MAR, we have
\begin{align}\nonumber
\pi({\cal Y}^{(mis)},{\cal X}_{1:T},\boldsymbol\psi|{\cal Y}^{(obs)},{\cal D})&=\frac{
\int\pi({\cal D}|{\cal Y}^{(obs)},\xi)\pi({\cal Y}^{(mis)},{\cal Y}^{(obs)},{\cal X}_{1:T},\boldsymbol\psi)\pi(\xi)d\xi}{
\int\pi({\cal D}|{\cal Y}^{(obs)},\xi)\pi({\cal Y}^{(obs)})\pi(\xi)d\xi}& \\ \nonumber
&=\pi({\cal Y}^{(mis)},{\cal X}_{1:T},\boldsymbol\psi|{\cal Y}^{(obs)}).
\end{align}

Handling the missing data is easy when using the MH within Gibbs sampling scheme of Section 3.  Using the observed data and the current values for the missing data, the full conditionals for ${\cal X}_{1:T}$ and $\boldsymbol\psi$ are unchanged.  The full conditional of ${\cal Y}^{(mis)}$ is, for any $y_{ijt}\in {\cal Y}^{(mis)}$, determined by $\pi(y_{ijt}=1|{\cal X}_{1:T},\boldsymbol\psi)=1/(1+\exp(-\eta_{ijt})),$ where $\eta_{ijt}$ is given in (\ref{obseq2}).  That is, including the missing data in the MH within Gibbs sampling amounts to an additional draw for each missing $y_{ijt}$ from a Bernoulli distribution with probability determined by (\ref{obseq2}).

\FloatBarrier
\section{PREDICTION}
\label{Prediction}
Predicting future links is an important and interesting problem.  Applications include recommender systems, terrorist networks, protein interaction networks, prediction of friendship networks, and more \citep{wang2007local,kashima2006parameterized,hopcroft2011will,liben2007link}.

When considering prediction in the latent space context, it is of interest to predict for time $T+1$ both the edges of the adjacency matrix $Y_{T+1}$ and the latent space positions ${\cal X}_{T+1}$.  It is simple to find point estimates of the latter since
\begin{eqnarray} \nonumber
\pi({\cal X}_{T+1}|Y_{1:T})
&=& \int \pi({\cal X}_{T+1}|{\cal X}_T,\boldsymbol\psi)\pi({\cal X}_{1:T},\boldsymbol\psi|Y_{1:T})d{\cal X}_{1:T}d\boldsymbol\psi  \\
&\approx& \frac{1}{L}\sum_{\ell=1}^L\prod_{i=1}^n N({\bf X}_{i(T+1)}| {\bf X}_{iT}^{(\ell)},{\sigma^2}^{(\ell)}I_p),
\label{expectedX}
\end{eqnarray}
where the superscript $^{(\ell)}$ indicates the $\ell^{th}$ draw from the posterior.  Hence
$\widehat{{\cal X}}_{T+1}:=\mathbb{E}({\cal X}_{T+1}|Y_{1:T})\approx \frac{1}{L}\sum_{\ell=1}^L {\cal X}_T^{(\ell)}$.  It is assumed that an appropriate burn-in period for the chain has been accounted for.

A simple way to compute a point estimate of the probability of an edge between $i$ and $j$ at time $T+1$, $\mathbb{P}(y_{ij(T+1)}=1)$, would be to plug in $\widehat{{\cal X}}_{T+1}$ along with the posterior means of the parameters into the observation equation (\ref{obseq2}).  We can, however, do a little better by not conditioning on the posterior means of the model parameters, hence eliminating some unnecessary uncertainty.  We aim, then, to find $\mathbb{P}(Y_{T+1}|Y_{1:T},{\cal X}_{T+1}=\widehat{{\cal X}}_{T+1})$.  Since conditional on ${\cal X}_{T+1}$ we still assume that the $y_{ij(T+1)}$'s are independent, we need only find $\mathbb{P}(y_{ij(T+1)}|Y_{1:T},\widehat{{\bf X}}_{i(T+1)},\widehat{{\bf X}}_{j(T+1)})$.  This can be estimated as follows:

First, we approximate the joint distribution.
\begin{align}\nonumber
&\mathbb{P}(y_{ij(T+1)},{\bf X}_{i(T+1)},{\bf X}_{j(T+1)}|Y_{1:T}) \\ \nonumber
=& \int \pi(y_{ij(T+1)}|{\bf X}_{i(T+1)},{\bf X}_{j(T+1)},\boldsymbol\psi) \pi({\bf X}_{i(T+1)},{\bf X}_{j(T+1)}|{\cal X}_{1:T},\boldsymbol\psi)\pi({\cal X}_{1:T},\boldsymbol\psi|Y_{1:T})d{\cal X}_{1:T}d\boldsymbol\psi \\
\approx& \frac{1}{L}\sum_{\ell=1}^L \pi(y_{ij(T+1)}|{\bf X}_{i(T+1)},{\bf X}_{j(T+1)},\boldsymbol\psi^{(\ell)}) \pi({\bf X}_{i(T+1)},{\bf X}_{j(T+1)}|{\cal X}_{T}^{(\ell)},\boldsymbol\psi^{(\ell)}).
\end{align}
Next the marginal distribution of $({\bf X}_{i(T+1)},{\bf X}_{j(T+1)})|Y_{1:T}$ is found as in (\ref{expectedX}) and approximated by
$\frac{1}{L}\sum_{\ell=1}^L N({\bf X}_{i(T+1)}|{\bf X}_{iT}^{(\ell)},{\sigma^2}^{(\ell)}I_p) N({\bf X}_{j(T+1)}|{\bf X}_{jT}^{(\ell)},{\sigma^2}^{(\ell)}I_p).$  Thus the conditional distribution of $y_{ij(T+1)}|Y_{1:T},\widehat{{\cal X}}_{T+1}$ is estimated as a weighted average:
\begin{eqnarray}\nonumber
\mathbb{P}(y_{ij(T+1)}|Y_{1:T},\widehat{{\cal X}}_{T+1}) &\approx&\sum_{\ell=1}^L w_{\ell}\pi(y_{ij(T+1)}|\widehat{{\bf X}}_{i(T+1)},\widehat{{\bf X}}_{j(T+1)},\boldsymbol\psi^{(\ell)}),
\label{predProbs}
\end{eqnarray}
where
\begin{equation}
w_{\ell}= \frac{N(\widehat{{\bf X}}_{i(T+1)}|{\bf X}_{iT}^{(\ell)},\boldsymbol\psi^{(\ell)})N(\widehat{{\bf X}}_{j(T+1)}|{\bf X}_{jT}^{(\ell)},\boldsymbol\psi^{(\ell)})}{\sum_{\ell'=1}^L N(\widehat{{\bf X}}_{i(T+1)}|{\bf X}_{iT}^{(\ell')},\boldsymbol\psi^{(\ell')})N(\widehat{{\bf X}}_{j(T+1)}|{\bf X}_{jT}^{(\ell')},\boldsymbol\psi^{(\ell')})}
\end{equation}
and $\pi(y_{ij(T+1)}|\widehat{{\bf X}}_{i(T+1)},\widehat{{\bf X}}_{j(T+1)},\boldsymbol\psi)$ is defined in (\ref{obseq1}) and (\ref{obseq2}).

This method outperforms the simpler plug in method mentioned earlier, as is shown in the Supplementary Material.  The intuition as to why this is so is that we are using fewer estimated parameters to make predictions, hence introducing less uncertainty into the prediction estimates.
\vspace{-1pc}

\FloatBarrier
\section{EDGE ATTRACTION}
\label{NodalInfluence}
Social influence is a well defined concept in the literature, which \cite{anagnostopoulos2008influence} defined as ``the phenomenon that the actions of a user can induce his/her friends to behave in a similar way."  Many authors attempt to use social influence to track the propagation of ideas or behaviors through a network (e.g., \cite{kempe2003maximizing}, \cite{leskovec2006patterns}).  \cite{tang2009social} described a method of determining which actors will influence which other actors on a variety of topics.  \cite{goyal2010learning} proposed a method of labeling each edge by an influence probability, assuming undirected binary edges.  These works all require data outside of the network, such as, as phrased by Goyal et al., an ``action log."  Here we consider a novel type of influence, called edge attraction, defined to be the attracting influence one actor has on another actor's friendships; e.g., in a social network this is how one person draws another person into their own social circle.  Hence our new type of influence is how one actor affects the edges of another actor.

We assume that the way in which one actor can affect another actor's movements in the network is manifested in an increased tendency for the influenced actor to move in the direction of the influencing actor in the social space.  To detect the tendency for an actor $i$ to move through the social space in the direction of another actor $j$, the transition equation for the latent actor positions is extended by considering a new parameter to describe the edge attraction between two actors.  We will then carefully define this parameter, implement an appropriate prior, and then look at posterior probabilities that will help the user determine whether or not there is edge attraction.  We will show that this can be done by using the same MCMC output as when the transition equation was assumed to be a random walk.  Throughout the next two sections we consider looking at the actors pairwise, i.e., we look at whether or not a specific actor $i$ is influenced by another actor $j$.

\subsection{Detection of Edge Attraction}
\begin{figure}[h!]
\centering
\includegraphics[scale=0.35]{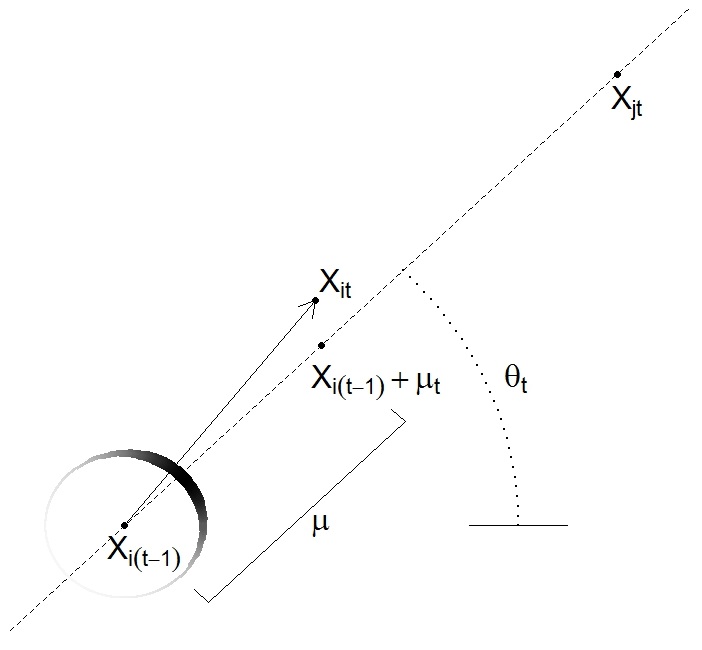}
\caption{The extension of the transition equation to allow for actor $j$'s influence on actor $i$.  Actor $i$ is more likely to move {\it toward} actor $j$.  The circle around ${\bf X}_{i(t-1)}$ represents a von Mises distribution for the angle component of $\boldsymbol\epsilon_{it}$'s polar coordinates, where dark values indicate high probability regions and light values indicate low probability regions.  }
\label{NI_fig2}
\end{figure}

Consider an extension of the transition equation (\ref{transition}) such that ${\bf X}_{it}={\bf X}_{i(t-1)}+\boldsymbol\epsilon_{it}$ where $\boldsymbol\epsilon_{it}\sim N(\boldsymbol\mu_t,\sigma^2I_p)$.  In the following, assume that $p=2$.  Let $\theta_t$ equal the angle $\mbox{atan2}({\bf X}_{jt}-{\bf X}_{i(t-1)})$, where atan2 is the common variation of the arctangent function which preserves the angle's quadrant, taking a vector rather than a ratio as its argument.  Let ${\cal R}_t$ be the rotation matrix associated with $\theta_t$.  We then let $\boldsymbol\mu_t$ be of the form
\vspace{-0.5pc}\begin{equation}
\boldsymbol\mu_t={\cal R}_t\left( \begin{array}{c} \mu \\ 0 \end{array} \right) = \left( \begin{array}{cc}
\cos(\theta_t)&-\sin(\theta_t)\\
\sin(\theta_t)&\cos(\theta_t)
\end{array} \right)\left( \begin{array}{c} \mu \\ 0 \end{array} \right),
\end{equation}
where $\mu=\|\mathbb{E}({\bf X}_{it}-{\bf X}_{i(t-1)})\|$ is some unknown parameter taking non-negative values.  No edge attraction is equivalent to the case where $\mu=0$, and if there does exist some edge attraction then this will be reflected in some $\mu>0$.  Figure \ref{NI_fig2} gives an illustration of this type of edge attraction.  The idea here is that actor $i$ will aim towards wherever actor $j$ is within the latent characteristic space.  If we let the prior on the parameters $(\boldsymbol\psi,\mu)$ be independent, i.e., $\pi(\boldsymbol\psi,\mu)=\pi(\boldsymbol\psi)\pi(\mu)$, then the posterior samples obtained from Section \ref{PosteriorSampling} can be equivalently viewed as having come from $\pi({\cal X}_{1:T},\boldsymbol\psi|Y_{1:T}, \mu=0)$.  This is important because, as will be seen later, we can use these same draws to make inference regarding the edge attraction existing between actors $i$ and $j$.  Also note that under the extended transition equation, the Markov property still holds for the latent positions, i.e., $\pi({\cal X}_t|{\cal X}_{1:(t-1)},\boldsymbol\psi,\mu)=\pi({\cal X}_t|{\cal X}_{t-1},\boldsymbol\psi,\mu)$ (see the Supplementary Material).

The prior distribution of $\mu$ is chosen to be a mixture of a point mass on 0 and a continuous component over the positive reals:
\begin{equation}
\pi(\mu)=\left\{\begin{array}{cl} p_0 & \mbox{if } \mu=0
\\ (1-p_0)f(\mu) & \mbox{for } \mu>0, \end{array} \right.
\end{equation}
where $f$ is some proper continuous density on $(0,\infty)$.  Here $f$ will be assumed for convenience to be the exponential distribution with mean $\lambda$.  Then the posterior density is
\begin{eqnarray}
\pi(\mu|Y_{1:T})=\frac{\pi(Y_{1:T}|\mu)\pi(\mu)}{\pi(Y_{1:T})}= \frac{\pi(Y_{1:T}|\mu)p_0}{\pi(Y_{1:T})}{\bf 1}_{\{\mu=0\}} + \frac{\pi(Y_{1:T}|\mu)}{\pi(Y_{1:T})}(1-p_0)f(\mu){\bf 1}_{\{\mu>0\}}.
\end{eqnarray}
For notation, let $\pi_0(\mu=0|Y_{1:T})=\pi(Y_{1:T}|\mu=0)p_0/\pi(Y_{1:T})$ and let $\pi_+(\mu|Y_{1:T})=\pi(Y_{1:T}|\mu)(1-p_0)f(\mu)/\pi(Y_{1:T})$.  Then $\pi_0(\mu=0|Y_{1:T})$ is the point mass posterior probability that $\mu=0$.  If our prior probability $p_0=1/2$ and we find that the posterior probability is less than 1/2 then this implies the data is pulling the posterior probability towards the conclusion that actor $i$ is influenced by actor $j$.

Since $1=\pi_0(\mu=0|Y_{1:T}) + \int_0^{\infty}\pi_+(\mu|Y_{1:T})d\mu$,
we have that
\begin{equation}
\pi_0(\mu=0|Y_{1:T})=\frac{1}{1+\int_0^{\infty}\kappa(\nu)d\nu},
\label{NIeq1}
\end{equation}
where $\kappa(\nu)=\pi_+(\mu=\nu|Y_{1:T})/\pi_0(\mu=0|Y_{1:T})$.  So if we can find $\int_0^{\infty}\kappa(\nu)d\nu$ then we can compute $\pi_0(\mu=0|Y_{1:T})$.  To this end, we have the following proposition whose proof is given in the Supplementary Material:
\begin{proposition}
For $\kappa(\nu)$ as defined above,
\begin{equation}
\int_0^{\infty}\kappa(\nu)d\nu = \mathbb{E}(h({\cal X}_{1:T},\boldsymbol\psi)|Y_{1:T},\mu=0),
\label{lemma1EQ1}
\end{equation}
where
\begin{equation*}
h(\cx_{1:T},\bPsi)= \frac{1-p_0}{\lambda p_0}\cdot\frac{\Phi(z)}{\sqrt{\frac{T-1}{\sigma^2}}\phi(z)}, \hspace{1pc}
z=\frac{\frac{1}{T-1}\sum_{t=2}^T\small(\bx_{it}-\bx_{i(t-1)})' \Big({\scriptsize\begin{array}{c}  \cos(\theta_t) \\ \sin(\theta_t) \end{array}} \Big)\normalsize - \frac{\sigma^2}{\lambda(T-1)}}{\sqrt{\sigma^2/(T-1)}},
\end{equation*}
$\boldsymbol\Phi$ is the standard normal cumulative distribution function, and $\phi$ is the standard normal density.
\label{lemma1}
\end{proposition}

The expectation in (\ref{lemma1EQ1}) is taken with respect to the posterior $\pi({\cal X}_{1:T},\boldsymbol\psi|Y_{1:T},\mu=0)$, and hence we can use the posterior draws already obtained from Section \ref{PosteriorSampling} to utilize the following approximation:
\begin{equation}
\int_0^{\infty}\kappa(\nu)d\nu \approx \frac{1}{N}\sum_{\ell =1}^N h({\cal X}_{1:T}^{(\ell)},\boldsymbol\psi^{(\ell)}).
\label{NIeq3}
\end{equation}
Combining (\ref{NIeq1}) with (\ref{NIeq3}) we are able to compute the posterior probability $\pi_0(\mu=0|Y_{1:T})$.

The quantity $z$ in Proposition \ref{lemma1} is interesting in that $({\bf X}_{it}-{\bf X}_{i(t-1)})' \scriptsize\Big(\begin{array}{c} \cos(\theta_t) \\ \sin(\theta_t) \end{array} \Big)\normalsize$ is the scalar projection of $({\bf X}_{it}-{\bf X}_{i(t-1)})$ onto the unit vector whose direction is determined by $({\bf X}_{jt}-{\bf X}_{i(t-1)})$.  Intuition tells us that if these scalar projections are consistently large then actor $j$ is influencing the way actor $i$ moves through the social space; the posterior probabilities reflect this intuition in that large scalar projections lead to small values of $\pi_0(\mu=0|Y_{1:T})$.

One last note of practical value is that for edge attraction to exist in a meaningful way, we must require that the influencing actor has during at least one observation period brought the influenced actor within his social circle.  This becomes easy to evaluate by means of the social reaches by requiring that for edge attraction to exist between $i$ and $j$, $\{t : d_{ijt}<r_i\}\bigcup\{t : d_{ijt}<r_j\} \neq \emptyset$.

\subsection{Visualizing the Edge Attraction}
\label{VisualizingtheNodalInfluence}
Suppose there is evidence from the posterior that $\mu\neq 0$.  Then, as mentioned earlier, $\boldsymbol\epsilon_{it}={\bf X}_{it}-{\bf X}_{i(t-1)}\sim N( \boldsymbol{\mu}_t, \sigma^2 I_p)$.  We can visualize and further interpret this influence by considering the polar coordinates of $\boldsymbol\epsilon_{it}$, $d_{it} =\|{\bf X}_{it}-{\bf X}_{i(t-1)}\|$ and $\phi_{it}=\mbox{atan2}({\bf X}_{it}-{\bf X}_{i(t-1)})$.  The following proposition, whose proof is given in the Supplementary Material, gives the distribution of $(d_{it},\phi_{it})$.
\begin{proposition}
Let $Z$ and $W$ be independent random variables such that $Z\sim N(\mu_z,\sigma^2)$ and $W\sim N(\mu_w,\sigma^2)$, and let $d$ $=\|(Z,W)\|$ and $\phi$ $=\mbox{atan2}((Z,W))$ be the polar coordinates of $(Z,W)$.  Then
\begin{equation}
d \sim \mbox{Rice}\left(\|(\mu_z,\mu_w)\|, \sigma \right), \hspace{2pc}
\phi|d \sim \mbox{von Mises}\left( \frac{d\|(\mu_z,\mu_w)\|}{\sigma^2}, \mbox{atan2}((\mu_z,\mu_w)) \right).
\end{equation}
\label{lemma2}
\end{proposition}
\vspace{-2.5pc}Using this proposition, we see that the polar coordinates $(d_{it}, \phi_{it})$ of ${\cal R}_t'\boldsymbol\epsilon_{it}$ follow
\begin{equation}
d_{it} \sim \mbox{Rice}\left(\mu,\sigma\right), \hspace{2pc}
\phi_{it} \mathlarger{\mathlarger{|}} d_{it} \sim \mbox{von Mises}\left( \frac{d_{it}\mu}{\sigma^2}, 0 \right).
\label{estimatedVM}
\end{equation}
In other words, we can think of the transition from ${\bf X}_{i(t-1)}$ to ${\bf X}_{it}$ as a two step process, where the distance to move is determined first, and then the angle is chosen.  To aid the visualization of the edge attraction we focus on the von Mises distribution that determines this angle.  {\it Hence we are visualizing the extent of the edge attraction from $j$ on $i$ by looking at the propensity of actor $i$ to aim towards actor $j$}.  The circle around ${\bf X}_{i(t-1)}$ in Figure \ref{NI_fig2} represents such a von Mises distribution (with mean $\theta_t$ rather than 0), where dark values indicate high probability regions and light values indicate low probability regions.  Note that if $\mu=0$ then $\phi_{it}\sim \mbox{von Mises}(0,0) \stackrel{{\cal D}}{=}\mbox{Unif}(-\pi,\pi)$.  That is, any angle with respect to actor $j$ is as likely as any other angle and hence actor $i$ does not tend to angle towards actor $j$ more than any other direction in the latent social space.

We can use the posterior mean latent positions to estimate these von Mises distributions, thus obtaining a good visualization of the edge attraction.  First get $\hat{\mu}$, the estimate of $\mu$, by averaging over time the scalar projection of $(\widehat{{\bf X}}_{it}-\widehat{{\bf X}}_{i(t-1)})$ onto $(\widehat{{\bf X}}_{jt}-\widehat{{\bf X}}_{i(t-1)})$.  Then we can further estimate the $T-1$ concentration parameters (the concentration parameter in (\ref{estimatedVM}) being $d_{it}\mu/\sigma^2$) by multiplying this $\hat{\mu}$ by $\|\widehat{{\bf X}}_{it}-\widehat{{\bf X}}_{i(t-1)}\|/\hat{\sigma}^2$.  One can then plot these estimated von Mises distributions wrapped around the actors being influenced, such as Figure \ref{knechtNI_AtoD}.  This type of plot may become overcrowded when there are multiple influencing actors; in such a case, for each of, say, $m$ influencing actors, one could average these $T-1$ concentration parameters over time to obtain one concentration parameter for a summary von Mises distribution.  These $m$ von Mises distributions can then be plotted to get an overview of the various influences on the influenced actor.  An example of this type of plot can be seen in the Supplementary Material.

\FloatBarrier
\section{SIMULATIONS}
\label{Simulations}
Twenty data sets were simulated, each with the number of actors $n=100$ and the number of time points $T=10$.  For each of the twenty simulations we set $\beta_{IN}=1$ and $\beta_{OUT}=2$, and randomly drew the radii $r_{1:n}$ from a Dirichlet distribution.  For ten of the twenty simulations, twenty-five actors were randomly selected to be influenced, each of which was accompanied by another randomly selected actor to do the influencing; there was no edge attraction incorporated in the remaining ten simulations.  Details on how the data were simulated, along with extra simulation results beyond those given below, can be found in the Supplementary Material.

The results from the simulations were compared in several ways:  First, for each simulation the posterior means of  $\beta_{IN}$ and $\beta_{OUT}$ were computed as well as the correlation between the posterior means of $r_{1:n}$ and the truth for each of the 20 simulations.  The mean (sd) over all 20 simulations of $\widehat\beta_{IN}$ was 0.9172 (0.06207) and for $\widehat\beta_{OUT}$ was 2.045 (0.1438).  The mean (sd) correlation between $\widehat r_{1:n}$ and $r_{1:n}^{(true)}$ was 0.9298 (0.06402).  We see then that the posterior means did quite well at estimating the true values of $\beta_{IN}$ (1), $\beta_{OUT}$ (2) and the radii.

\begin{figure}[h!]
\centering
\subfigure[AUC]
{
\includegraphics[height=6cm]{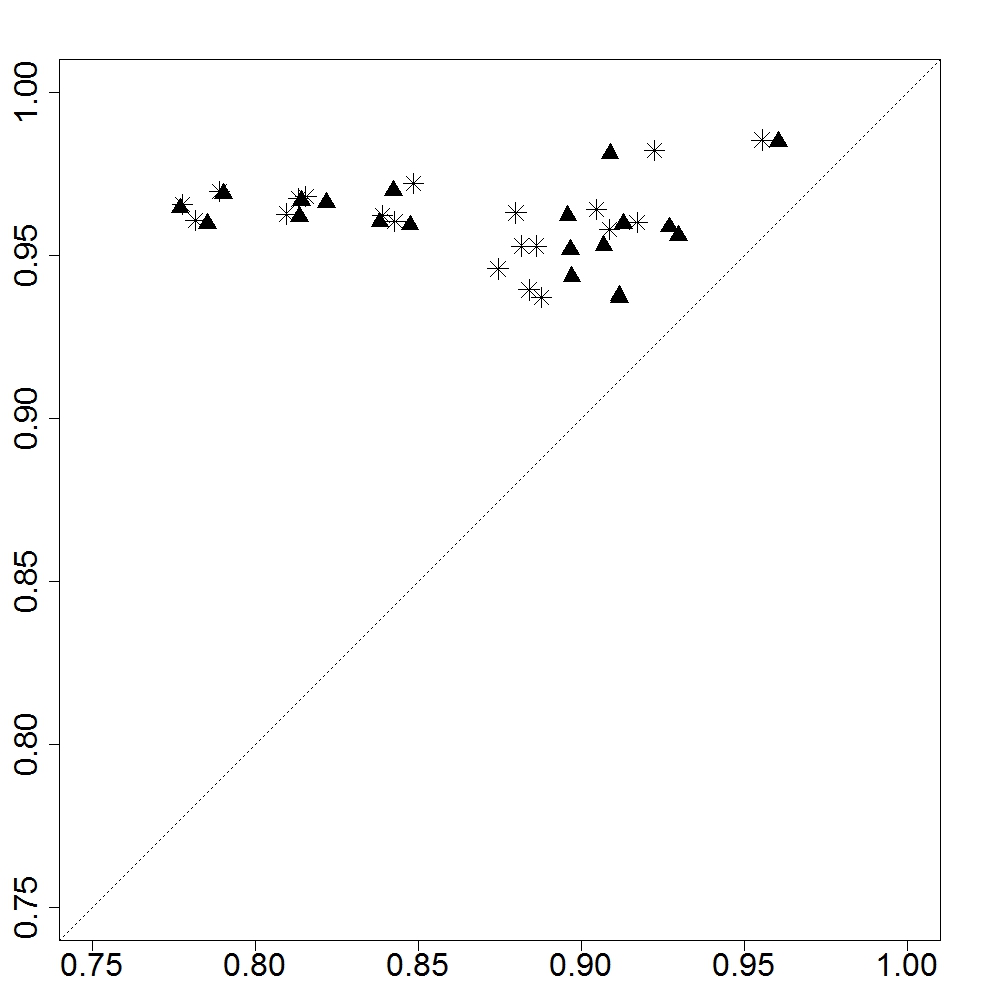}
\label{simResults_a}
}
\subfigure[Distances]
{
\includegraphics[height=6cm]{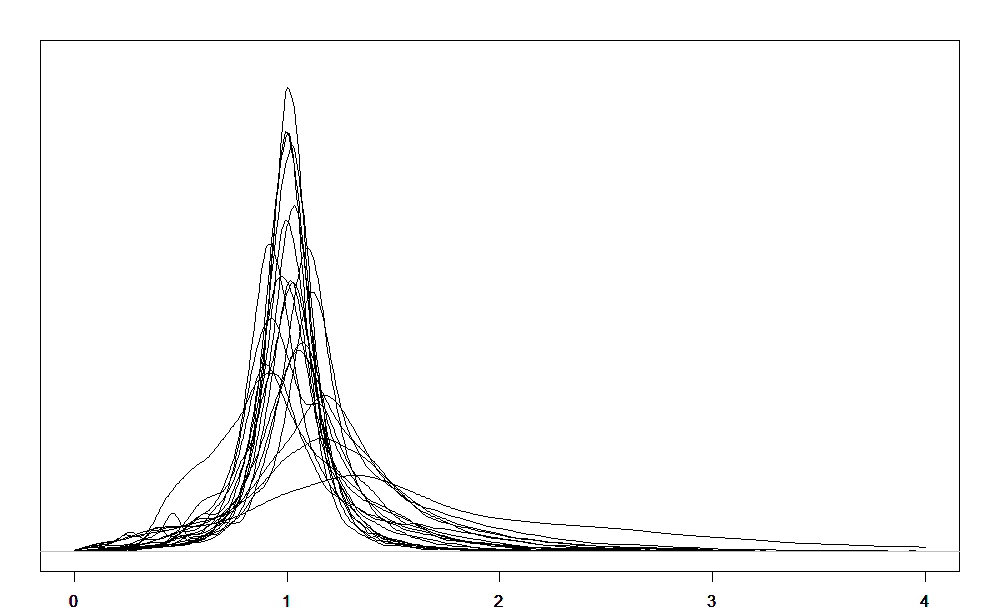}
\label{simResults_b}
}
\caption{Results for 20 simulations. (a) AUC using Sarkar and Moore's method (horizontal axis) and our method (vertical axis) on both undirected (triangles) and directed (asterisks) networks; (b) Distribution of pairwise distance ratios, comparing estimated latent positions with true latent positions.}
\end{figure}

Second, the area under the ROC curve (AUC) was computed.  This was accomplished by plugging in the posterior means of the model parameters and the latent positions into the observation equations (\ref{obseq1}) and (\ref{obseq2}), and then comparing these with the simulated data $Y_{1:T}$.  Hence this can be considered as a measure of how well the model fits the data.  For each simulation, the directed graphs were also converted to undirected graphs by letting $y_{ijt}=\max\{y_{ijt},y_{jit}\}$ in order that we might apply the method of \cite{sarkar2005dynamic}.  The AUC values for both undirected and directed networks were then computed using the estimates from Sarkar and Moore's method.  We similarly computed the AUC values for the directed network using our method, and, again using those same estimates, computed the AUC values for the undirected network by using $\mathbb{P}(\{y_{ijt}=1\}\cup \{y_{jit}=1\})$.  These results are given in Figure \ref{simResults_a}.  We see that all our values are extremely high, implying that the model fits the data quite well, and also we see that our method uniformly outperformed that of Sarkar and Moore on both the directed and undirected networks.

Third, the pairwise distances from the estimated latent positions were compared to the pairwise distances from the true latent positions.  That is, for each triple $(i,j,t)$ we can look at $\|\widehat{{\cal X}}_{it}-\widehat{{\cal X}}_{jt}\|/\|{\cal X}_{it}-{\cal X}_{jt}\|$, giving us $Tn(n-1)/2$ such ratios for each simulation.  Figure \ref{simResults_b} gives, for each of the twenty simulations, a smoothed curve of the distribution of these ratios.  Notice that all these distributions are narrow and centered around 1, implying that the latent positions from the posterior means are close to the truth.

For those ten cases where edge attraction was part of the simulation, we computed the sensitivity of detecting edge attraction on those actors which were in truth influenced, and in all 20 simulations we computed the specificity of not detecting influence on those actors which were in truth not influenced.  The mean (sd) sensitivity and specificity for the ten simulations with edge attraction were 0.952 (0.0316) and 0.832 (0.129).  The mean (sd) specificity for the ten simulations without edge attraction was 0.868 (0.116).  We see from this that the Bayesian estimation does a very good job at detecting edge attraction without giving many false positives when no such influence exists.

The Supplementary Material provides a discussion on sensible priors for all model parameters except $\sigma^2$.  It is important then to determine the sensitivity of the MCMC algorithm to the values of $\theta_{\sigma}$ and $\phi_{\sigma}$.  To this end we reran the above simulations where the shape and scale parameters of $\pi(\sigma^2)$ were drawn from a uniform distribution ranging from 3 to 15 for the shape parameter and from 0.01 to 2 for the scale parameter.  The AUC for rerunning these 20 simulations in this fashion yielded very high AUC values, ranging from 0.9407 to 0.9858, averaging 0.9621.  Thus it appears that the estimation is quite robust to the hyperparameters for the prior of $\sigma^2$.

In addition to the simulations described above, five larger data sets were simulated where $n=500$ and $T=10$.  Estimation was performed both using and not using the approximations outlined in Section \ref{Scalability} and in the Supplementary Material, letting $n_0=100$, and the AUC was computed to evaluate model fit.  Simulations were analyzed on a UNIX machine with a 2.40 GHz processor.  The mean (sd) time to perform the MCMC analysis with 50,000 iterations using the approximation was, in minutes, 716 (24), and to perform the MCMC with 50,000 iterations not using the approximation was, in minutes, 2281 (20).  Hence by using the approximations there was a mean (sd) decrease in computational time of 68.6\% (0.835).  The mean (sd) AUC using the approximation was 0.9618 (0.0048), and not using the approximation was 0.9679 (0.0109).  Thus by using the approximations of Section \ref{Scalability} there is a drastic decrease in computational time with very little loss in model fit.

\section{REAL DATA ANALYSIS}
\label{RealDataAnalyses}
\subsection{Dutch Classroom Data}
\begin{figure}[t]
\centering
\includegraphics[width=\textwidth]{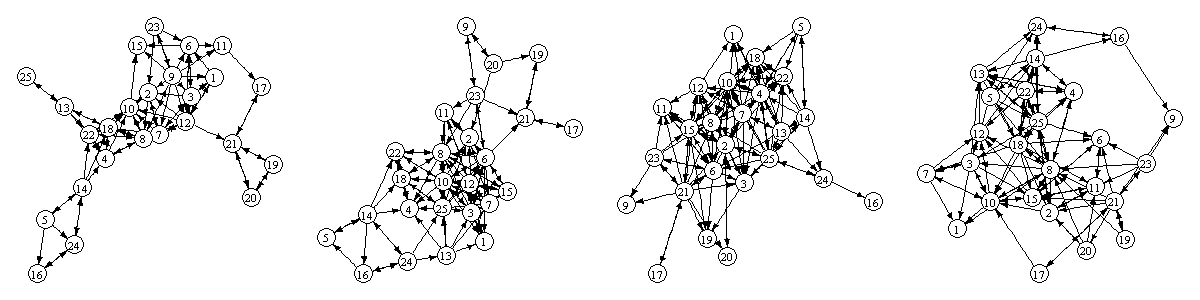}
\caption{Graphs of Dutch classroom data at, from left to right, times 1, 2, 3, and 4.}
\label{knecht_data}
\end{figure}

\cite{knecht2008friendship} conducted a longitudinal study in which students aged 11 to 13 years in a Dutch class were surveyed over four time points, yielding four asymmetric adjacency matrices where the $(i,j)^{th}$ entry denotes whether student $i$ claims student $j$ as a friend.  Figure \ref{knecht_data} shows the graphs from these adjacency matrices.  Demographic and behavioral data were also collected on these individuals.  Twenty six students were recorded, although one student left the class before the study was completed; this student was left out of the analysis.  Missing edges exist in the data due to some students not being present during a survey.  This was dealt with as previously described in Section \ref{MissingData}.

The trace plots of $\beta_{IN}$, $\beta_{OUT}$, $\sigma^2$, and $\tau^2$ are given in the Supplementary Material.  A burn-in of 15,000 iterations was removed, leaving a chain of length 85,000.
We compared our method with that found in \cite{sarkar2005dynamic} by AUC values.  Our method yielded an AUC value of 0.917 vs. 0.8456 from Sarkar and Moore's method.

\begin{figure}[h!]
\centering
\includegraphics[width=0.7\linewidth]{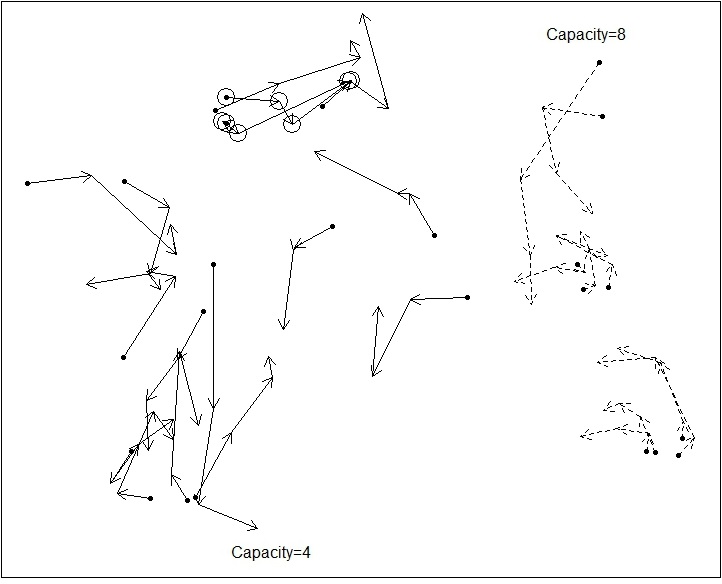}
\caption{Posterior means of latent actor positions for the Dutch classroom data, arrows indicating the temporal direction of the trajectories.  Males' trajectories are in dotted lines, females' in solid lines. Also, two of the students are of non-Dutch ethnicity, and these two actors' latent positions are circled.  The two students with the lowest (4) and highest (8) ranked academic capabilities, actors 9 and 25 respectively, are also marked as such.}
\label{knechtDems}
\end{figure}

The posterior means of $\beta_{IN}$ and $\beta_{OUT}$ were 1.29 and 1.00 respectively, implying that popularity was more important in edge formation.  The posterior means of the latent locations are given in Figure \ref{knechtDems}.  Some interesting features can be noticed by comparing the latent positions with demographic information.  Figure \ref{knechtDems} differentiates the actors' gender by males as dotted lines and females as solid lines.  This plot corroborates results shown by \cite{snijders2010introduction} in that the friendships between two actors of the same gender are more prevalent.  Also, only two of the students are of non-Dutch ethnicity, circled in Figure \ref{knechtDems}, and these two actors are very close together in the social space.  One last interesting feature seen in Figure \ref{knechtDems} regards an interesting link between academic capability and social behavior.  Each student was assessed and ranked from 4 (lowest) to 8 (highest) based on academic capabilities at the end of primary school.  There was only one student (actor 9) who was ranked a 4 and one (actor 25) who was ranked an 8.  The social behavior of these two individuals, as seen via the latent space positions, are complete opposites.  The student with the highest ranked capability moves from outside the social network to the center of the social space, while the student with the lowest ranked capabilities moves directly away from the center of the social space.  The reason actor 25 started outside of the network may be explained in part by the fact that he had only gone to primary school with one other student and hence did not start the school term knowing the others.

\begin{figure}[h!]
\centering
\includegraphics[height=6cm]{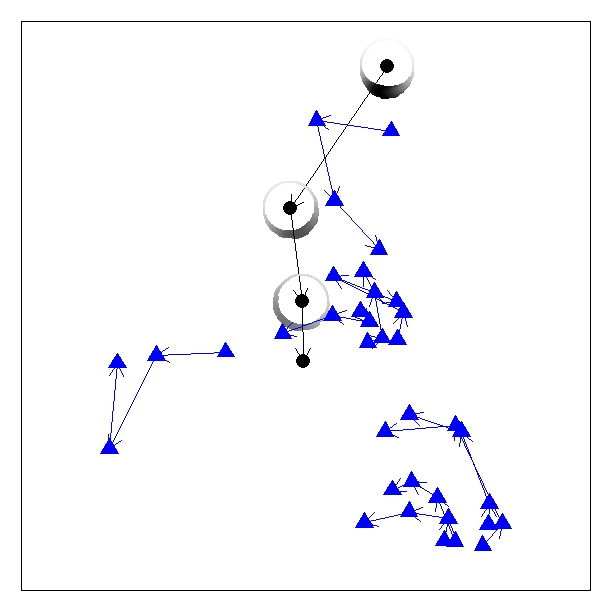}
\caption{Corresponding to the Dutch classroom data, this plot is zooming in on the posterior means of the latent positions of the influenced actor (circle), student 25, and those actors doing the influencing (triangles).  The circles around the influenced actor are the von Mises distributions from (\ref{estimatedVM}) that help to visualize the influence being exerted in terms of the direction the influenced actor moves.  Wider and darker areas on the rings indicate higher probability regions.}
\label{knechtNI_AtoD}
\end{figure}

Edge attraction was detected in four of the actors.  Of note is the fourth such influenced actor (actor 25), who was unique in that he was influenced by many of the other actors (9 others).
Figure \ref{knechtNI_AtoD} gives the posterior mean of the latent positions of these actors with a wrapped von Mises distribution plotted around actor 25 corresponding to the strongest influencer (largest $\mu$).  Plots of the von Mises distributions for each of the individual influencing actors are given in the Supplementary Material.  From Figure \ref{knechtNI_AtoD} we can visualize the strength and direction of the influence, thus yielding detailed information on the local scale of the network.
As mentioned earlier, actor 25 began by only knowing one other actor (determined by having or not having gone to the same primary school as the others), and so it is intuitive that he would be more susceptible to being pulled into others' existing social circles, bringing him to the center of the social space.  The von Mises distribution in Figure \ref{knechtNI_AtoD} matches this intuition in that the edge attraction wanes as time progresses and actor 25 becomes a part of his own social circle.  Finally, the results from the analysis of edge attraction fell in line with the overall gender and ethnic separation, in that the first three instances of edge attraction all occurred within the ethnic Dutch girls, and of the nine actors influencing actor 25 (a male) only one was of the opposite gender.


\subsection{Cosponsorship Data}
\label{Cosponsorship}
We analyzed data collected by James Fowler on bill cosponsorship of Congressmen in the U.S. House of Representatives for the $97^{th}$ to $101^{st}$ Congresses (see \cite{fowler2006connecting} and \cite{fowler2006legislative}).  There were a total of 644 members of Congress (MC's) who served during these five terms.  However, at each time point around 30\% of MC's were not represented (the actual values ranged from 30.1\% to 31.1\%).  These large proportions of unrepresented MC's leads to even larger proportions of missing edges (from 51.2\% to 52.5\% missing edges).  The data were analyzed by letting $y_{ijt}=1$ if actor $j$ sponsored a bill and actor $i$ cosponsored it, hence showing support for actor $j$.  The graphs from these adjacency matrices from time 1 (97$^{th}$ Congress) to 5 (101$^{st}$ Congress), sans missing data, are given in the Supplementary Material.

Due to the large amount of missing data, some care was needed in initializing the missing edges in the Markov chain.  We modified the preferential attachment method of imputation described by \cite{huisman2008treatment} by doing the following.  We first form an aggregated adjacency matrix $Y$ whose entries $y_{ij}$ are set to one if for any $t$ there is a link from $i$ to $j$ and set to zero otherwise.  Next, for each missing actor $i$ (at a particular time point $t$), we assign the probability of a link from $i$ to $j$ to be proportional to the indegree (averaged over $t$) of actor $j$ and inversely proportional to the shortest path length between $i$ and $j$ in $Y$.  That is, letting $k_j$ denote the average indegree of actor $j$ and $n_{ij}$ denote the shortest path length between $i$ and $j$ in $Y$, the probability of a link from $i$ to $j$ is set to be
$\mathbb{P}(y_{ijt}=1)=(k_j/n_{ij})/(\sum_{\ell\neq i}k_{\ell}/n_{i\ell})$.  We then look at the average outdegree of $i$ (rounded to the nearest integer), denoted $d_i$, and randomly draw $d_i$ actors from $\{1,\ldots,n\}\setminus\{i\}$ using these probabilities; the corresponding $y_{ijt}$'s are set to be 1.  In this way we obtain initial values for the missing data.

The trace plots of $\beta_{IN}$, $\beta_{OUT}$, $\sigma^2$, and $\tau^2$ are given in the Supplementary Material.  A burn-in of 250,000 iterations was removed, leaving a chain of length 1,250,000.  Thinning was done by recording only every tenth iteration.  Using the posterior means to make predictions on $Y_{1:T}$ led to an AUC value of 0.787 vs. 0.7148 from Sarkar and Moore's method; when applying Sarkar and Moore's method we used $Y_{1:T}$ constructed from the observed edges and imputed edges.  From these AUC values we see that our model fits the data quite well and again outperforms the existing method.

Many of the congressmen (328 MC's) analyzed were reelected into the 102$^{nd}$ Congress, and so it was possible to compare predictions with the truth.  In addition to the predictions obtained through the methods described in Section \ref{Prediction}, we also considered prediction by using $\sum_{t=1}^Ty_{ijt}/T$ to estimate $\mathbb{P}(y_{ij(T+1)}=1)$.  For both averaging $Y_{1:T}$ and applying our method, hard predictions were made by letting $\widehat{y}_{ij(T+1)}=1$ if $\widehat{\mathbb{P}}(y_{ij(T+1)}=1)>0.5$ and 0 otherwise.  Table \ref{PredictionTable} gives the results.  From this we see that while using a more na\"{i}ve prediction method yields higher specificity, it does so at the expense of correctly detecting the future edges.  Our method can better find the future edges, and also gives the lowest mean squared error (MSE).
\begin{table}[htb]
\centering
\begin{tabular}{c|r|r|r}
Method & \multicolumn{1}{c|}{Specificity} & \multicolumn{1}{c|}{Sensitivity} & \multicolumn{1}{c}{MSE} \\ \hline \hline
Averaging $Y_{1:T}$ & 0.8014 & 0.5125& 0.2492 \\ \hline
$\mathbb{P}(Y_{T+1}|Y_{1:T},\widehat{{\cal X}}_{T+1})$ & 0.5962 & 0.6984 & 0.2151
\end{tabular}
\caption{Prediction results for 328 MC's in our analysis who also served in the $102^{nd}$ Congress.}
\label{PredictionTable}
\end{table}

\begin{figure}[htb]
\centerline{\includegraphics[scale=0.325]{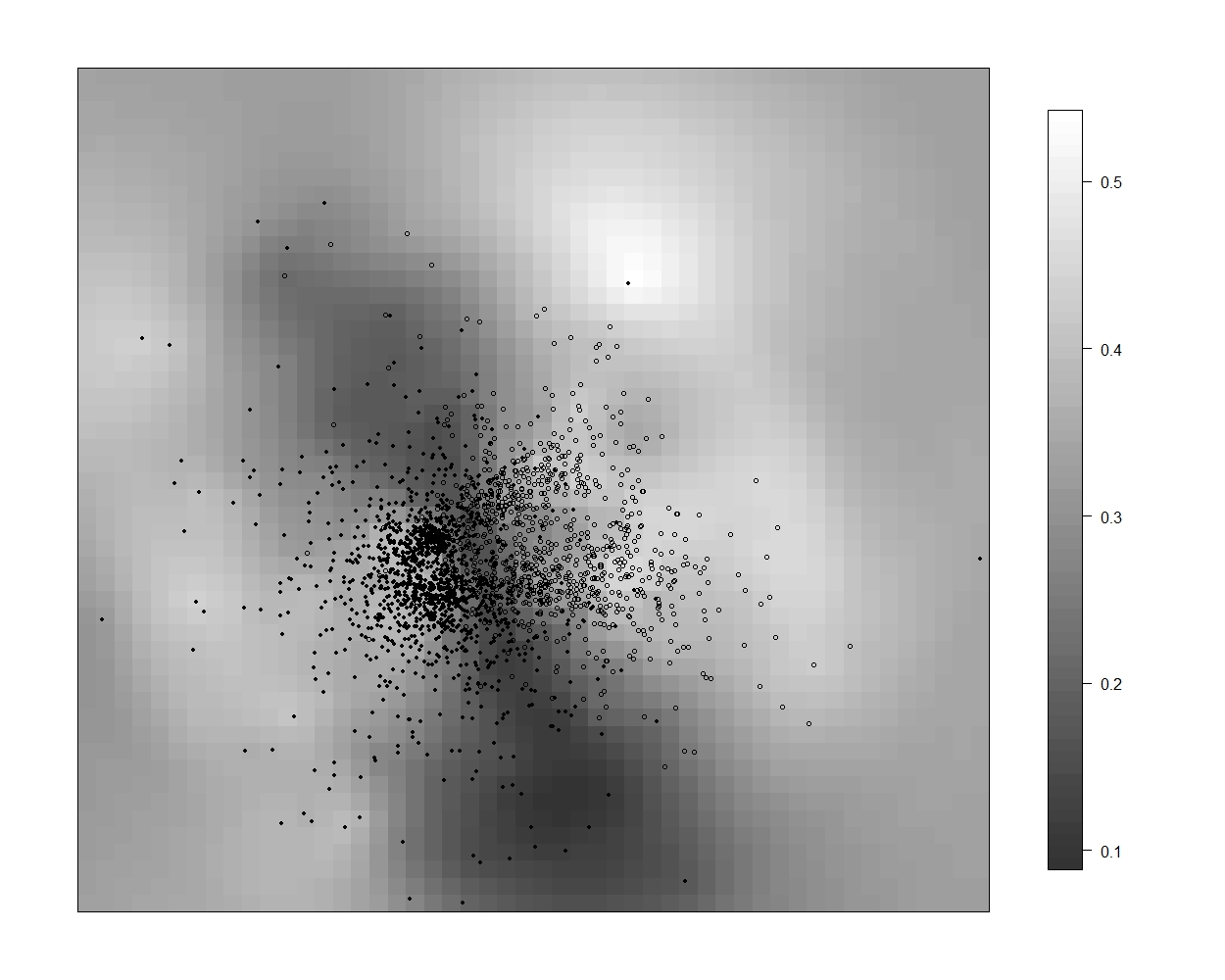}\vspace{-2pc}}
\caption{Posterior means of latent positions for the cosponsorship data.  Latent positions for all 5 Congresses are plotted simultaneously.  Hollow circles are republicans and solid circles are democrats.  The surface these points lie upon reflects the political ideological landscape within the network space.  Darker regions correspond to more moderate ideologies, and lighter regions correspond to more radical ideologies.}
\label{cosp_ideology}
\end{figure}

The posterior means of the coefficients, $\beta_{IN}=0.974, \beta_{OUT}=0.147$, indicate that popularity was dramatically more responsible for creating edges than activity level; i.e., the probability of a cosponsorship is determined mostly by the MC who sponsors the bill rather than the MC who is contemplating cosponsoring it.  Figure \ref{cosp_ideology} shows the posterior mean latent positions of the MC's.  Unsurprisingly we see the Republicans and Democrats occupy different halves of the network space.  Both parties seem to have the majority of their members in the center of the network space along with a scattering of members along the edge of the network space, implying that both parties have active central members which associate with members from both parties, as well as less active outlying members which interact less with members of the opposite party.

\begin{figure}[htb]
\centering
\includegraphics[height=5cm]{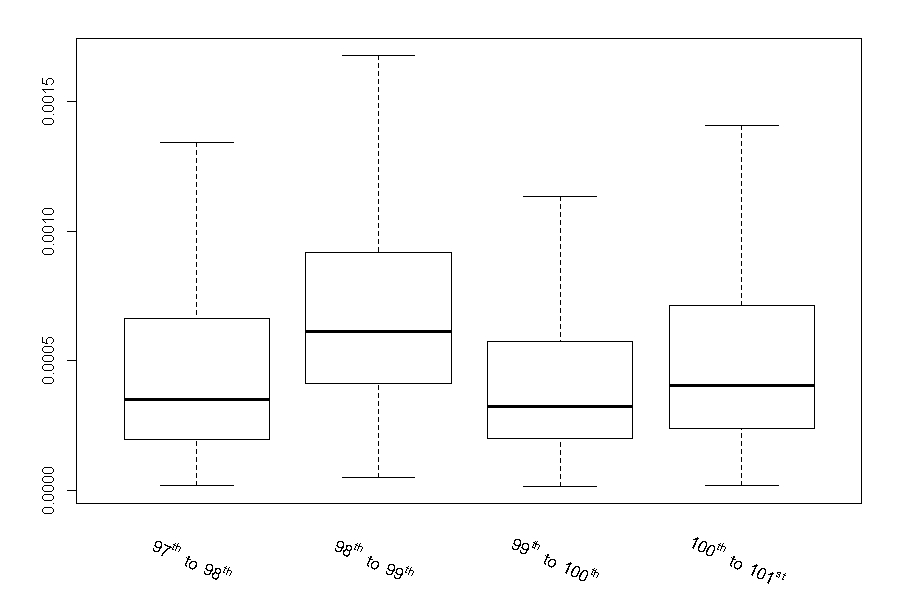}
\caption{Boxplots of the distances MC's traveled within the latent network space during each of the four transitions.  The similar ranges imply that the dynamics of the network are fairly constant throughout the five terms.}
\label{cosp_dynamicsBoxplot}
\end{figure}

It is of interest to study the dynamics of the network.  To evaluate the stability of the network we consider the distance each MC moves during each of the four transitions.  Figure \ref{cosp_dynamicsBoxplot} gives a boxplot for these distances, and from this we can see that the distances corresponding to each transition fall within a similar range, though the transition to the $99^{th}$ Congress involves somewhat larger moves.  There were a few MC's (ranging from 4.4\% to 7.7\% of the MC's) who were above the top whisker, but these typically were different MC's at every transition; only 11 of the MC's were beyond the top whisker in two of the transitions, 2 of the MC's in three of the transitions, and none more than three.  All this indicates that the dynamics of the network remained stable throughout the five terms.

Political ideology, measured from liberal to conservative, is an extremely important aspect of political science.  Much literature exists on this topic; for example, \cite{poole2011ideology} wrote an entire book on ideology and its effect on Congress. \cite{levitt1996senators} discussed various factors' effects on roll-call voting patterns, concluding that personal ideology is the single most important factor.  This relationship between voting patterns and personal ideology is seen in a vivid way by comparing the latent positions of the MC's with their ideologies.  Specifically, this comparison is shown in Figure \ref{cosp_ideology}, where the latent positions of the MC's are superimposed upon a surface which represents a political ideological landscape.  This surface was obtained in the following way.  Each MC has a particular Nominate score which is a measure of their political ideology \citep[see][]{poole1985spatial}.  This score ranges from $-1$ (liberal) to 1 (conservative).  Using the latent location coordinates, kriging was performed on the absolute value of the Nominate scores using a spherical variogram model.  This gives us the surface in Figure \ref{cosp_ideology} that reflects how regions of the latent network space correspond to radical ideologies or moderate ideologies.
In the center of the network space where the actors are most dense (and hence are more active in legislation) is an interesting dividing line between the two parties that reflects a moderate political ideology.  We also see that both parties have a less dense (hence less active in legislation) group of MC's which has more radical ideologies.

Edge attraction was detected on 74 of the MC's.  It is intuitive that an MC would be influenced more by members of his or her own party than by members of a different party, and indeed this is the case.  Of the influenced MC's, only 29\% were influenced more by members of the opposite party than by members of their own party.  As an example of an MC influenced by members of the opposite party, consider Lawrence Coughlin, a Republican from Pennsylvania.  Only 35\% of those who exerted influence on Coughlin were also Republicans, and in fact the average Nominate score for those exerting influence on Coughlin was $-0.073$, i.e., Coughlin was influenced mostly by slightly liberal politicians.  This influence is manifest in the fact that he is often referred to as a moderate Republican \citep[e.g.,][]{caughlin2001obit}; his moderate ideology (0.163) is also quantitatively reflected in having his Nominate score below the first quartile of fellow Republicans, and below the first quartile of the absolute value of the Nominate scores of all MC's.  In contrast to Coughlin, consider Sidney Yates, a Democrat from Illinois.  94\% of those exerting influence on Yates were also Democrats, and in fact quite liberal Democrats;  the mean ideology score of Yates' influencing MC's was $-0.301$ (recall that a negative Nominate score implies liberal ideology).  The influence of these liberal MC's on Yates is reflected in Yates also being liberal, himself having a Nominate score ($-0.477$) below the first quartile of all Democrats and an absolute score above the third quartile of the absolute values of all Nominate scores.  What is left uncertain is whether Yates aimed towards liberal Democrats in the latent space because he himself already had a liberal ideology or whether these liberal Democrats influenced him to become liberal himself.  The von Mises distributions corresponding to the edge attraction on both Coughlin and Yates are given in the Supplementary Material.

\section{DISCUSSION}
\label{Discussion}
A latent space model is given for analyzing dynamic network data. The model provides rich visualization of the dynamics of the network, giving insight into the characteristics of the actors, the evolution of the network, and the overall groupings and communities that exist within the network. Unlike existing methodology, our model can handle directed edges, missing data, and can be used to predict future latent positions and future edges, and detect and visualize edge attraction.  We have also given an approximation method to obtain statistically meaningful estimates in a computationally efficient way.

While only directed graphs have been analyzed, our methods can easily be used to model undirected graphs.  Clearly with undirected edges there is no distinction between activity and popularity.  This can be reflected in the model by setting $\beta_{IN}$ = $\beta_{OUT}$ in equation (\ref{obseq2}) and proceeding as before.

While the focus of this paper is binary edges, this model can be easily generalized to dyadic data types other than binary.  This is accomplished by changing the link function just as one would in the generalized linear model setting.  In (\ref{obseq2}) we see that the link function $\eta$ up to this point has been assumed to be the logit of the conditional mean of $y_{ijt}$, $\mathbb{E}(y_{ijt}|{\cal X}_t,\boldsymbol\psi)$.  If, for example, we were dealing with dyadic relations measured in counts, then it may be better to let $\eta$ be the log of $\mathbb{E}(y_{ijt}|{\cal X}_t,\boldsymbol\psi)$, the canonical link for a Poisson random variable.   A similar approach has been taken in the static case by \cite{hoff2005bilinear}, and in the dynamic case by \cite{sewell2015analysis} for rank-order data; this latter work was completed after the present paper, building on the methodology proposed here.  Other data types may lead to similar adaptations of the model.

{\small

\bibliographystyle{asa}

\bibliography{JASA_flsm_refs}
}

\end{document}